# Effective Field and the Bloch–Siegert Shift at Bichromatic Excitation of Multiphoton EPR

A. P. Saĭko[a], G. G. Fedoruk[b], and S. A. Markevich[a]

[a] *Joint Institute of Solid State and Semiconductor Physics, National Academy of Sciences of Belarus, Minsk, 220072 Belarus*
e-mail: saiko@ifttp.bas-net.by
[b] *Institute of Physics, University of Szczecin, 70-451 Szczecin, Poland*

The dynamics of multiphoton transitions in a two-level spin system excited by transverse microwave and longitudinal RF fields with the frequencies $\omega_{mw}$ and $\omega_{rf}$, respectively, is analyzed. The effective time-independent Hamiltonian describing the "dressed" spin states of the "spin + bichromatic field" system is obtained by using the Krylov–Bogoliubov–Mitropolsky averaging method. The direct detection of the time behavior of the spin system by the method of nonstationary nutations makes it possible to identify the multiphoton transitions for resonances $\omega_0 = \omega_{mw} + r\omega_{rf}$ ($\omega_0$ is the central frequency of the EPR line, $r = 1, 2$), to measure the amplitudes of the effective fields of these transitions, and to determine the features generated by the inhomogeneous broadening of the EPR line. It is shown that the Bloch–Siegert shifts for multiphoton resonances at the inhomogeneous broadening of spectral lines reduce only the nutation amplitude but do not change their frequencies.

PACS numbers: 32.80.Wr, 33.35.+r, 42.50.Gy

In view of the possibilities for changing the energy states of the quantum state with external electromagnetic fields and its interaction with the environment, interest in studying the dynamics of multiphoton processes in nonlinear optics, NMR and EPR, has increased recently [1–8]. In two-level spin systems, there are no real intermediate levels, and "dressed" spin systems formed by external electromagnetic fields can serve as intermediate levels. In view of this circumstance, a multilevel dynamical quantum system with controllable properties can be created on the basis of the two-level system. Analysis of the dynamics of the dressed spin systems opens new perspectives for developing methods in order to narrow spectral lines, to separate overlapping spectra, and to increase the existence time of coherent spin states.

In pulsed EPR experiments based on the simultaneous absorption or emission of several photons with noticeably different frequencies, e.g., microwave and RF radiation, two- and three-photon electron spin echo [1], two-photon nutation [7], as well as RF-field-induced transparency [3] and nutation on dressed states [4, 5], were observed. Owing to the small amplitude of the effective field (effective Rabi frequency) of multiphoton transitions, as well as to the presence of inhomogeneous line broadening in real systems, difficulties remain both in direct observation of the dynamics of these transitions and in its quantitative interpretation on the basis of adequate theoretical approaches.

In this work, the dynamics of multiphoton transitions in two-level spin systems with its excitation by bichromatic (transverse microwave and longitudinal RF) radiation is studied theoretically and experimentally. To this end, the nonstationary nutation phenomenon, which is present the dynamics of quantum transitions, is used. The nutation frequency is directly determined by the effective interaction field. The direct observation of the dynamics of the interaction of radiation with the spin system makes it possible to reveal regions of multiphoton transitions caused by the inhomogeneous broadening.

In the semiclassical approximation, the Hamiltonian of the electron spin system (spin $S = 1/2$) in the linearly polarized microwave field directed along the $x$ axis of the laboratory coordinate system, as well as in the linearly polarized RF and static magnetic fields directed along the $z$ axis, can be written in the form

$$H(t) = \omega_0 s^z + 2\omega_1 \cos(\omega_{mw} t + \varphi) s^x + 2\omega_2 \cos(\omega_{rf} t + \psi) s^z, \quad (1)$$

where $\omega_0 = \gamma B_0$ is the resonance frequency of spin variables in the magnetic field $B_0$, $\gamma$ is the electron gyromagnetic ratio, $\omega_1 = \gamma B_1$ and $\omega_2 = \gamma B_2$ are the Rabi frequency, and $B_1$, $B_2$, $\omega_{mw}$, $\omega_{rf}$, $\varphi$, and $\psi$ are the amplitudes, frequencies, and phases of the microwave and RF fields, respectively.

The time evolution of the spin system is described by the density operator ρ, which satisfies the Liouville equation

$$i\frac{\partial \rho}{\partial t} = [H(t), \rho] \quad (2)$$

(Planck's constant $\hbar = 1$).

In the presence of the small parameter $\omega_1/\omega_{rf}$ (or $\omega_2/\omega_{rf}$), the evolution of the spin system with the time-dependent Hamiltonian given by Eq. (1) can be analyzed in the framework of perturbation theory with the use of the averaging procedure over oscillations developed by Krylov, Bogoliubov, and Mitropolsky [9].

In the case under consideration, it is convenient to pass to a new coordinate system by means of the canonical transformation

$$\rho \rightarrow \rho' = U^+ \rho U,$$
$$U = \exp\left\{-i2\int^t d\tau \omega_2 \cos(\omega_{rf}\tau + \psi) s^z\right\}. \quad (3)$$

As a result, Eq. (2) has the form

$$i\frac{\partial \rho'}{\partial t} = [H(t), \rho'] \equiv [H_0 + H_1(t), \rho']. \quad (4)$$

Here,

$$H_0 = \omega_0 s^z, \quad (5)$$

$$H_1(t) = \omega_1 e^{-i(\omega_{mw}t + \varphi)} e^{i\frac{2\omega_2}{\omega_{rf}}\sin(\omega_{rf}t + \psi)} s^+ + \text{H.c.}$$
$$= \omega_1 \sum_{n=-\infty}^{\infty} J_n\left(\frac{2\omega_2}{\omega_{rf}}\right) e^{-i(\omega_{mw} - n\omega_{rf})t} e^{-i(\varphi - n\psi)} s^+ + \text{H.c.}; \quad (6)$$

where $J_n(x)$ is the Bessel function. When deriving Eq. (6), we neglect "antirotating" terms in the operator $\cos(\omega_{mw}t + \varphi)s^x$ of the initial Hamiltonian. The Krylov–Bogoliubov–Mitropolsky method in the canonical formalism [9] (see also [10]) is implemented by excluding fast oscillating terms in the Hamiltonian $H_1(t)$ in any order of perturbation theory in the parameter $\omega_1/\omega_{rf}$ (whose smallness is assumed). This leads to the time-dependent effective Hamiltonian of the system and simplifies the problem of determining the evolution of dynamical variables.

If the quantum system is described by the Hamiltonian $H = H_0 + H_1(t)$, where $H_0$ is the unperturbed "diagonal" part of the energy operator and $H_1(t)$ is the small perturbation, "off-diagonal" term, in the interaction representation

$$\tilde{H}_1(t) = e^{iH_0 t} H_1(t) e^{-iH_0 t}, \quad (7)$$

the fast motion with the period $2\pi/\|H_0\|$ is imposed on a slower evolution of the system, which is characterized by the time $\|H_1\|^{-1}$, because $\|H_0\| \gg \|H_1\|$ by hypothesis (here, $\|H\|$ means the magnitude of the operator $H$ in frequency units). The application of the averaging procedure [9, 10] to the equation $\sigma = e^{iH_0 t} \rho' e^{-iH_0 t}$ for the density matrix $i\partial\sigma/\partial t = [\tilde{H}_1(t), \sigma]$ eliminates fast oscillating terms in it. As a result, we arrive at the equation

$$i\partial\langle\sigma\rangle/\partial t = [H_{eff}, \langle\sigma\rangle] \quad (8)$$

for the averaged density matrix $\langle\sigma\rangle$, which presents the slow evolution of the system in times $\|H_1\|^{-1}$ with the approximately "diagonal" effective Hamiltonian $H_{eff}$ (the averaging operation is denoted as

$$\langle A(t)\rangle = \frac{1}{T}\int_0^T A(t) dt,$$

where $T$ is the fast oscillation period, which is equal to or a multiple of $2\pi/\|H_0\|$). In particular, up to the third order in interaction, the effective Hamiltonian has the form

$$H_{eff} = H_{eff}^{(1)} + H_{eff}^{(2)} + H_{eff}^{(3)}, \quad (9)$$

$$H_{eff}^{(1)} = \langle\tilde{H}_1(t)\rangle, \quad (10)$$

$$H_{eff}^{(2)} = \frac{i}{2}\left\langle\left[\int^t d\tau(\tilde{H}_1(\tau) - \langle\tilde{H}_1(\tau)\rangle), \tilde{H}_1(t)\right]\right\rangle, \quad (11)$$

$$H_{eff}^{(3)} = -\frac{1}{3}\left\langle\left[\int^t d\tau(\tilde{H}_1(\tau) - \langle\tilde{H}_1(\tau)\rangle), \left[\int^t d\tau\right.\right.\right.$$
$$\left.\left.\left.\times(\tilde{H}_1(\tau) - \langle\tilde{H}_1(\tau)\rangle), \left(\tilde{H}_1(t) + \frac{1}{2}\langle\tilde{H}_1(t)\rangle\right)\right]\right]\right\rangle, \quad (12)$$

where [,] means commutator.

Thus, the evolution of the system is determined by the Hamiltonian $H_{eff}$ that does not include a fast time dependence and is calculated by Eqs. (7)–(12) using Eqs. (5) and (6) for $H_0$ and $H_1(t)$, respectively.

We determine $H_{eff}$ under the resonance condition

$$\omega_0 - \omega_{mw} - r\omega_{rf} + \delta = 0, \quad (13)$$

where $r$ is an integer (positive or negative) number of absorbed or emitted photons of the RF field and $\delta$ is a certain detuning from the resonance such that $\delta \ll \omega_{rf}$ and $\omega_{mw}, \omega_0 \gg \omega_{rf}$. Substituting Eqs. (5) and (6) into Eqs. (7) and (10)–(12), we obtain

$$H_{eff}(r, \delta t) = H_{eff}^{(1)}(r, \delta t) + H_{eff}^{(2)}(r) + H_{eff}^{(3)}(r, \delta t),$$
$$H_{eff}^{(1)}(r, \delta t) = \omega_1^{(1)}(r) e^{-i\delta t} e^{-i(\varphi + r\psi)} s^+ + \text{H.c.}, \quad (14)$$

3          SAĬKO et al.

$$H_{\text{eff}}^{(2)}(r) = \Delta_{BS}(r)s^z, \quad (15)$$

$$H_{\text{eff}}^{(3)}(r, \delta t) = \omega_1^{(3)}(r)e^{-i\delta t}e^{-i(\varphi + r\psi)}s^+ + \text{H.c.}, \quad (16)$$

$$\omega_1^{(1)}(r) = \omega_1 J_{-r}(z), \quad (17)$$

$$\omega_1^{(3)}(r) = -\frac{2}{3}\omega_1^3 \Bigg\{ \sum_{n, n' \neq -r} \frac{J_n(z)J_{n'}(z)}{((r+n)\omega_{\text{rf}} - \delta)} \\
\times \frac{(J_{n+n'+r}(z) + J_{n'-n-r}(z))}{((r+n')\omega_{\text{rf}} - \delta)} \\
- \frac{1}{2}\sum_{n \neq -r} \frac{J_n(z)J_{-n-2r}(z)J_{-r}(z)}{(r+n)^2\omega_{\text{rf}}^2 - \delta^2} \\
+ \frac{1}{2}\sum_{n \neq -r} \frac{J_n^2(z)J_{-r}(z)}{((r+n)\omega_{\text{rf}} - \delta)^2} \Bigg\}, \quad (18)$$

$$\Delta_{BS}(r) = \sum_{n \neq r} \frac{2\omega_1^2}{(r-n)\omega_{\text{rf}} - \delta} J_n^2(z). \quad (19)$$

The argument of all Bessel functions is given by the expression $z = 2\omega_2/\omega_{\text{rf}}$, and the argument $\delta t$ in $H_{\text{eff}}(r, \delta t)$ means a slow time dependence due to the presence of the "slow" factors $e^{\pm i\delta t}$. These factors can be removed by passing to the coordinate system rotating with the frequency $\delta$ entering in Eq. (13):

$$H_{\text{eff}}(r, \delta t) \longrightarrow H_{\text{eff}}(r) = (\omega_0 + \Delta_{\text{BS}}(r) \\
- \omega_{\text{mw}} - r\omega_{\text{rf}})s^z + H_{\text{eff}}^{(1)}(r) + H_{\text{eff}}^{(3)}(r), \quad (20)$$

where $H_{\text{eff}}^{(1),(3)}(r) \equiv H_{\text{eff}}^{(1),(3)}(r, \delta t = 0)$.

According to Eqs. (14), (16), and (20), the amplitude of the effective field up to the third order in the parameter $\omega_1/\omega_{\text{rf}}$ is equal to

$$\omega_{\text{eff}}(r) = \omega_1^{(1)}(r) + \omega_1^{(3)}(r), \quad (21)$$

where $\omega_1^{(1)}(r)$ and $\omega_1^{(3)}(r)$ are defined by Eqs. (17) and (18), respectively.

The Hamiltonian $H_{\text{eff}}^{(1)}(r)$ describes the processes of the absorption (emission) of one quantum of the microwave field and $r$ photons of the RF field, and these processes are accompanied by multiple photon transitions, because the expression for the effective Rabi frequency (effective amplitude of the microwave field) includes the Bessel function of the argument $2\omega_2/\omega_{\text{rf}}$. For example, the absorption of one quantum of the microwave field and $r$ photons of the RF field can be accompanied by the virtual absorption (emission) of $m$ quanta and emission (absorption) of $m$ quanta of the microwave field and $r$ photons of the RF field ($m$ is an arbitrary integer). The term $H_{\text{eff}}^{(3)}(r)$ is a third-order correction to the effective field amplitude in $\omega_1/\omega_{\text{rf}}$ and thereby is noticeable for not too small $\omega_1/\omega_{\text{rf}}$ values. The quantity $\Delta_{BS}(r)$ in $H_{\text{eff}}^{(2)}(r)$ is the shift of the resonance frequency of the multiphoton transition, i.e., the Bloch–Siegert shift, which is caused by the interaction of the spin system with the nonresonant harmonics of the RF field. According to Eq. (19), the shift vanishes for $r = \delta = 0$.

Using Eq. (20) and taking into account Eq. (21), we arrive at the following expression for the absorption component of the nutation signal (see, e.g., [11]):

$$v(t) \sim \frac{\omega_{\text{eff}}(r)}{\beta(\Delta, r)} \exp\left\{-\frac{1}{T_2}\left(1 - \frac{1}{2}\frac{\omega_{\text{eff}}^2(r)}{\beta^2(\Delta, r)}\right)t\right\} \\
\times \sin(2\pi\beta(\Delta, r)t); \quad (22)$$

$$\beta(\Delta, r) = \sqrt{\omega_{\text{eff}}^2(r) + \delta^2(\Delta, r)}, \\
\delta(\Delta, r) = -\delta + \Delta_{\text{BS}}(r) + \Delta, \quad (23)$$

where $\delta(\Delta, r)$ is the detuning from resonance, which consists of the detuning from the line center $\delta$ stochastic component $\Delta$, and $\Delta_{BS}(r)$, and $T_2$ is the spin–spin relaxation time. For slow fluctuations, $\Delta$ describes the inhomogeneous broadening of the spin transition. The nutation signal given by Eq. (22) must be averaged with respect to $\Delta$ with the weight function of the Gaussian form $g(\Delta) = (T_2^*/\sqrt{\pi})\exp(-\Delta^2 T_2^{*2})$:

$$\langle v(t) \rangle = \int_{-\infty}^{\infty} d\Delta g(\Delta) v(t), \quad (24)$$

where $T_2^*$ is the time of the reversible phase relaxation. In this case, the dependence on $\Delta$ (through $\delta$) in Eqs. (17)–(19) can be neglected.

Nonstationary nutation excited by the bichromatic radiation under consideration was studied on a specially designed pulse spectrometer of the 3-cm EPR range [7]. In this case, continuous microwave and RF fields are used, and the resonance is established by means of a pulse of the longitudinal magnetic field. The equilibrium spin system is initially in the nonresonant static magnetic field $B = B_0 - \Delta B$. Then, the magnetic field at the time $t = 0$ changes stepwise to the value $B_0$, implementing the multiphoton resonance. Owing to the Zeeman effect associated with the jump in the magnetic field $\Delta B$, the quantum transition frequency of the spin system changes to the value $\omega_0$ and becomes equal to the sum of the microwave and RF fields. For the time interval of the pulse action, the resonant interaction of the total microwave and RF fields with the spin system was established. The resonance conditions are established in about 120 ns, which is much less than the spin–lattice ($T_1$) and spin–spin ($T_2$) relaxation times, ensuring the excitation of multiphoton nutation.

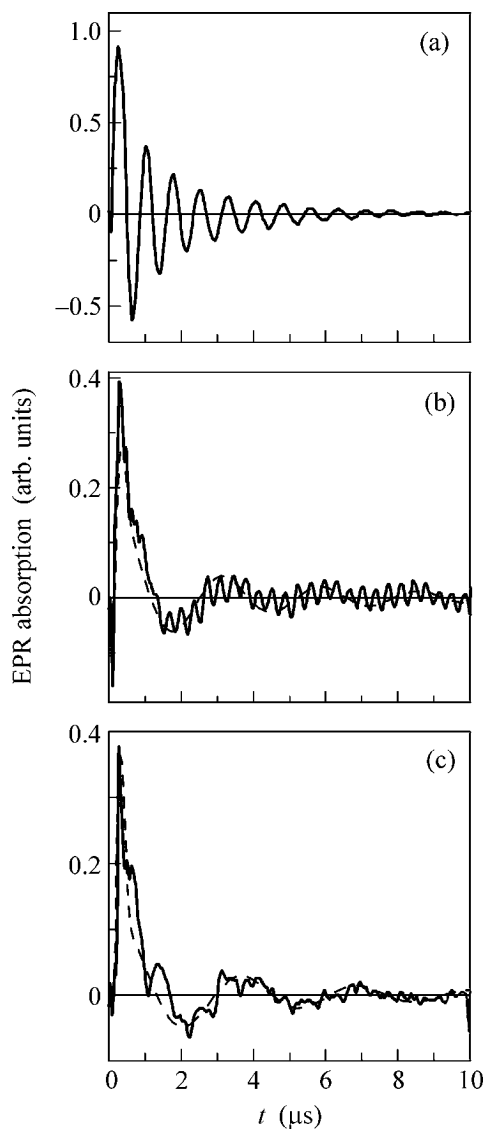

**Fig. 1.** Nutation EPR signals of $E'_1$ centers detected for the microwave field amplitude $\omega_1/2\pi = 1.33$ MHz: (a) the one-photon ($\omega_2 = 0$) nutation and multiphoton nutation for (b) two-photon ($r = 1$, $\omega_{rf} = 2.8$ MHz, $z = 0.57$) and (c) three-photon ($r = 2$, $\omega_{rf} = 1.4$ MHz, $z = 1.50$) resonances.

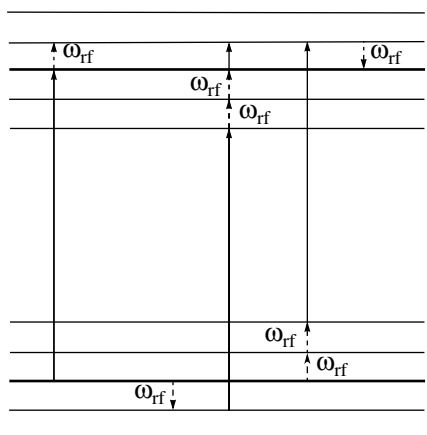

**Fig. 2.** Simplest multiphoton transitions excited at $\omega_0 = \omega_{mw} + \omega_{rf}$.

The RF field and magnetic field pulse were created by passing current through the same modulation element, which was located inside the measuring resonator. The amplitudes of microwave and RF fields were determined with an accuracy of about 5% and (in frequency units) did not exceed 1.5 MHz. To improve the signal-to-noise ratio, the multichannel digital summation of signals was used. The phase of the RF field was not matched to the onset of the magnetic field pulse.

The experiments were carried out at room temperature. As the two-level spin system, $E'_1$ centers in neutron-irradiated crystalline quartz were used [7]. The static magnetic field was parallel to the optical axis of the crystal. The EPR spectrum of $E'_1$ centers for a given orientation consists of a single line with the width $\Delta B = 0.016$ mT. In this case, the duration of magnetic field pulses was equal to 10 μs, their amplitude was $\Delta B = 0.12$ mT, and pulse repetition period was 1.25 ms.

Figure 1 shows the absorption signals of EPR nutations of the $E'_1$ centers detected for the fixed microwave field amplitude ($\omega_1/2\pi = 1.33$ MHz) for the one-, two-, and three-photon resonances. The signal in Fig. 1a detected at $\omega_{mw} = \omega_0$ and $\omega_2 = 0$ is associated with the usual one-photon nutation. The multiphoton nutation involving virtual RF photons that is observed at two-photon resonance $\omega_0 = \omega_{mw} + \omega_{rf}$ ($r = 1$, $\omega_{rf} = 2.8$ MHz, $z = 0.57$) is shown in the oscillogram in Fig. 1b. Figure 2 schematically shows the simple multiphoton transitions excited in this case. The oscillogram in Fig. 1c illustrates the multiphoton nutation observed at the three-photon resonance $\omega_0 = \omega_{mw} + 2\omega_{rf}$ ($r = 2$, $\omega_{rf} = 1.4$ MHz, $z = 1.50$).

The detected signal was the average of a large (up to $10^3$) number of signals obtained in the random RF field. This made it possible to decrease the corresponding oscillations of signals at the frequencies $\omega_{rf}$ and $2\omega_{rf}$ and to improve the detection of multiphoton nutation.

In view of the inhomogeneous broadening of the EPR line of the $E'_1$ centers, the frequencies of observed nutations are almost independent of the detuning $\delta$ from the resonance. The frequency of one-photon nutation in the absence of the RF field is equal to the Rabi frequency $\omega_1 = \gamma B_1$. The dashed lines in Figs. 1b and 1c are the results of the calculation by Eq. (24) for $T_2 = 4$ μs and $T_2^* = 1$ μs, respectively. The calculated dependences agree well with the observed signals. In this case, the frequencies of multiphoton nutations at two- and three-photon resonances within the measurement error correspond to the effective-field amplitudes calculated by Eqs. (17), (18), and (21).

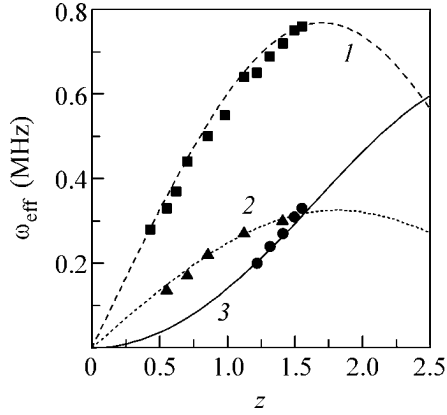

**Fig. 3.** Frequencies of multiphoton nutations excited at two- and three-photon resonances vs. the parameter $z = 2\omega_2/\omega_{rf}$ for (*1*) $r = 1$, $\omega_1/2\pi = 1.33$ MHz; (*2*) $r = 1$, $\omega_1/2\pi = 0.56$ MHz; and (*3*) $r = 2$, $\omega_1/2\pi = 1.33$ MHz.

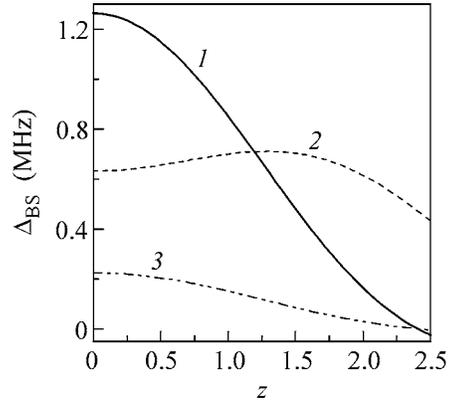

**Fig. 4.** Bloch–Siegert shifts vs. $z$ at two- and three-photon resonances for $\omega_{rf}/2\pi = 2.8$ MHz for (*1*) $r = 1$, $\omega_1/2\pi = 1.33$ MHz; (*2*) $r = 2$, $\omega_1/2\pi = 1.33$ MHz; and (*3*) $r = 1$, $\omega_1/2\pi = 0.56$ MHz.

Figure 3 shows the frequencies of multiphoton nutations excited at ($r = 1$) two- and ($r = 2$) three-photon resonances that are calculated and measured as a function of the parameter $z = 2\omega_2/\omega_{rf}$ (normalized RF field amplitude). The RF field frequency varies in the range $\omega_{rf} = 1.35$–3.5 MHz.

The presence of the Bloch–Siegert frequency shifts for the multiphoton resonances is directly manifested in the time behavior of the nutation signals (Fig. 4). This manifestation depends qualitatively on the character of the broadening of resonance lines. Indeed, for a homogeneously broadened line, the amplitude of the nutations decreases and their frequency increases (compared to the case where the shift is absent), whereas, for an inhomogeneously broadened line, only the amplitude decreases at unchanged nutation frequency (Fig. 5). This behavior is due to the fact that the Bloch–Siegert shift for inhomogeneous broadening detunes only a part of spin packets from resonance. At the same time, all spin packets are detuned from resonance at the homogeneous broadening and, according to Eq. (23), the nutation frequency increases and amplitude decreases.

Thus, the multiphoton transitions at two- and three-photon resonances of the bichromatic field with the two-level spin system are directly observed using non-stationary nutation. The use of the Krylov–Bogoliubov–Mitropolsky averaging method allowed calculation of the effective field amplitudes and the Bloch–Siegert frequency shifts directly used with the frequency and amplitude of the observed nutation signals. It has been found that the Bloch–Siegert shift for the inhomogeneously broadened line reduces only the nutation amplitude but does not change the nutation

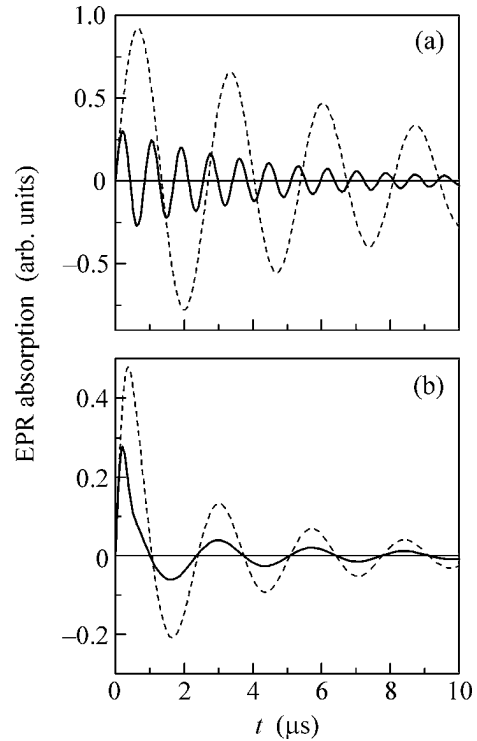

**Fig. 5.** Nutation signals for (a) homogeneous and (b) inhomogeneous broadening of the lines for $r = 1$, $\omega_1/2\pi = 1.33$ MHz, $\omega_{rf}/2\pi = 2.8$ MHz, and $z = 0.57$. The dashed lines show the signals in the absence of the Bloch–Siegert shifts.

frequencies, in contrast to the homogeneous broadening case.

## REFERENCES

1. I. Gromov and A. Schweiger, J. Magn. Res. **146**, 110 (2000).
2. M. Fedin, M. Kälin, I. Gromov, and A. Schweiger, J. Chem. Phys. **120**, 1361 (2003).
3. M. Kälin, I. Gromov, and A. Schweiger, Phys. Rev. A **69**, 033809 (2004).
4. G. Jeschke, Chem. Phys. Lett. **301**, 524 (1999).
5. G. G. Fedoruk, Fiz. Tverd. Tela (St. Petersburg) **46**, 1581 (2004) [Phys. Solid State **46**, 1631 (2004)].
6. P. T. Eles and C. A. Michal, J. Chem. Phys. **121**, 10167 (2004).
7. G. G. Fedoruk, Zh. Éksp. Teor. Fiz. **127**, 1216 (2005) [JETP **100**, 1069 (2005)].
8. H. Hatanaka, M. Sugiyama, and N. Tabuchi, J. Magn. Res. **165**, 293 (2003).
9. N. N. Bogoliubov and Yu. A. Mitropolsky, *Asymptotic Methods in the Theory of Nonlinear Oscillations*, 4th ed. (Nauka, Moscow, 1974; Gordon and Breach, New York, 1962).
10. A. P. Saĭko, Fiz. Tverd. Tela (St. Petersburg) **35**, 38 (1993) [Phys. Solid State **35**, 20 (1993)].
11. É. A. Manykin and V. V. Samartsev, *Optical Echo-Spectroscopy* (Nauka, Moscow, 1984) [in Russian].